\documentclass{PoS}

\title{Monte Carlo approach to the string/M-theory}

\ShortTitle{Monte Carlo approach to the string/M-theory}

\author{\speaker{Masanori Hanada}%
    \\
        KEK Theory Center\\
        E-mail: \email{hanada@post.kek.jp}}

\abstract{
It has long been conjectured that certain supersymmetric Yang-Mills (SYM) theories provide us with nonperturbative formulations of the string/M-theory. Although the supersymmetry (SUSY) on lattice is notoriously difficult in general, for a class of theories important for the string/M-theory various lattice and non-lattice methods, which enable us to study them on computers, have been proposed by now. In this talk, firstly I explain how SYM and string/M-theory are related. Then I explain why the lattice SUSY is difficult in general, and how the difficulties are solved in theories related to string/M-theory. Then I review the status of the simulations. It is explained that some stringy effects are correctly incorporated in SYM.\ Furthermore, concrete values can be obtained from the SYM side, even when a direct calculation on the string theory side is impossible by the state-of-the-art techniques. We also comment on other recent developments, including the  membrane mini-revolution in 2008 and simulation of the matrix model formulation of the string theory.

}

\FullConference{The 30th International Symposium on Lattice Field Theory\\
                 June 24 -- 29,  2012\\
                 Cairns, Australia}

\begin{document}

\section{Introduction}
The second superstring revolution in mid 1990's made it clear that the lattice Monte-Carlo simulation can play 
crucial roles in order to understand important problems associated with nonperturbative nature of the superstring/M theory, 
e.g. the microscopic origin of the black hole thermodynamics, the Big Bang singularity, the inflation, and the compactification of the extra dimensions. 
The key is a deep connection between certain supersymmetric gauge theories and superstring/M theory: 
it had been conjectured that the gauge theories provide the nonperturbative formulations of superstring/M-theory  
\cite{Banks:1996vh,Ishibashi:1996xs,MatrixString,Maldacena:1997re,Itzhaki:1998dd}. 
According to these conjectures, one can study the nonperturbative aspects of string/M-theory by solving the gauge theories. 
In particular, the gauge/gravity duality conjecture \cite{Maldacena:1997re,Itzhaki:1998dd} 
relates the thermodynamics of the super Yang-Mills theory and the black hole. 

The first excitement had declined rather quickly, partly because the numerical techniques to study the supersymmetric gauge theories were rather premature.  
However now various tools are available, some interesting physics have been revealed, and the studies of other important problems are within reach. 
In this talk I briefly review basics of this subject. I start with giving a very elementary introduction of the gauge/gravity duality in Sec.~\ref{sec:gauge_gravity_duality}.  
Then in Sec.~\ref{sec:SYM} I review the numerical approach. In Sec.~\ref{sec:ABJM} I review the relationship between the supersymmetric Chern-Simons-matter theory 
and M-theory on the multiple M2-brane background. In Sec.~\ref{sec:others} I show an incomplete list of other attempts. 

Because of the page limit, my explanation will be rather brief, and I cannot cover all the important topics. 
I recommend the readers to read other reviews as well, e.g. \cite{Catterall:2009it,Nishimura:2012xs,Catterall:2010nd}.

\section{The gauge/gravity duality}\label{sec:gauge_gravity_duality}
\subsection{Open string, closed string and supergravity}
\begin{figure}[htbp] 
\begin{center}
\scalebox{0.4}{\includegraphics{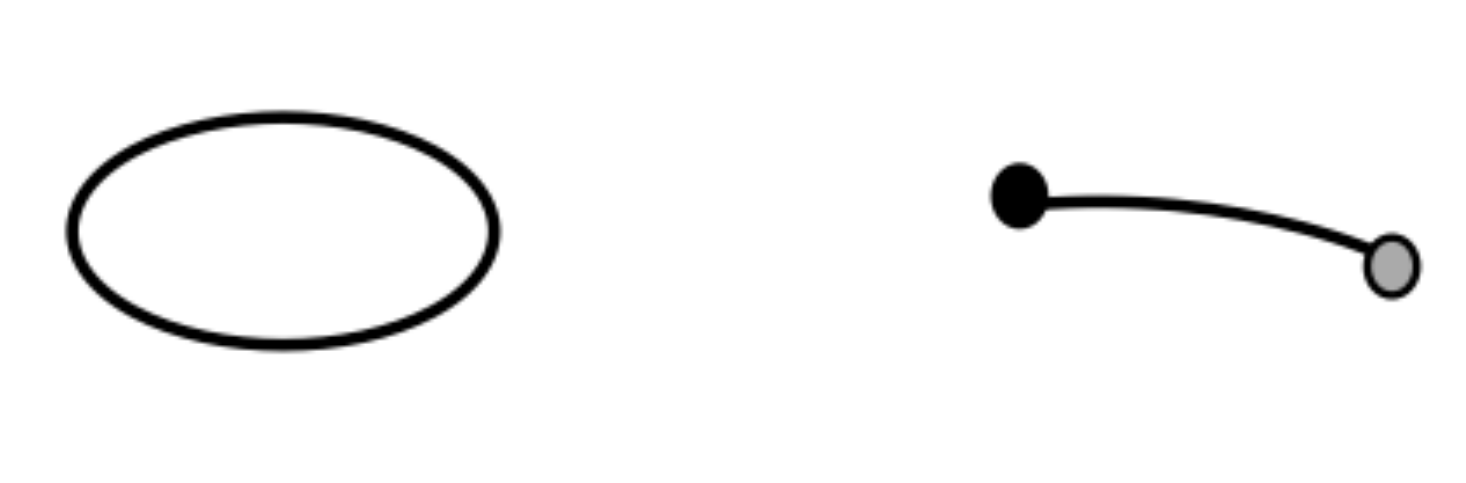}}
\caption{ 
A closed string (left) and an open string (right). 
Roughly speaking, the closed string describes graviton, 
while the open string corresponds to the gauge field.  
}\label{fig:strings}
\end{center} 
\end{figure}

In the superstring theory, point particles are promoted to strings. Each oscillatory mode of the strings can be regarded as a particle, 
whose mass is heavier when the oscillation is stronger. There are two kinds of strings, a closed string and an open string (Fig.~\ref{fig:strings}). 
The massless sector of the closed string describes the graviton, a scalar field called dilaton, and various tensor fields. 
The open string can have gauge degrees of freedom at the edges, and hence it naturally describes the gauge fields. 
Other fields e.g. the massless fermions and massive fields of the Planck scale mass also exist in the spectrum. 
The string perturbation theory describe the interaction of these fields. For consistency, $(1+9)$-dimensional spacetime 
(1 time dimension and 9 spatial dimensions) is required, and in order to relate it to our $(1+3)$-dimensional spacetime extra six spatial dimensions 
must become invisible, for example through the compactification. In this talk I consider the noncompact $(1+9)$-dimensional spacetime unless otherwise stated.


\begin{figure}[htbp] 
\begin{center}
\scalebox{0.25}{\includegraphics{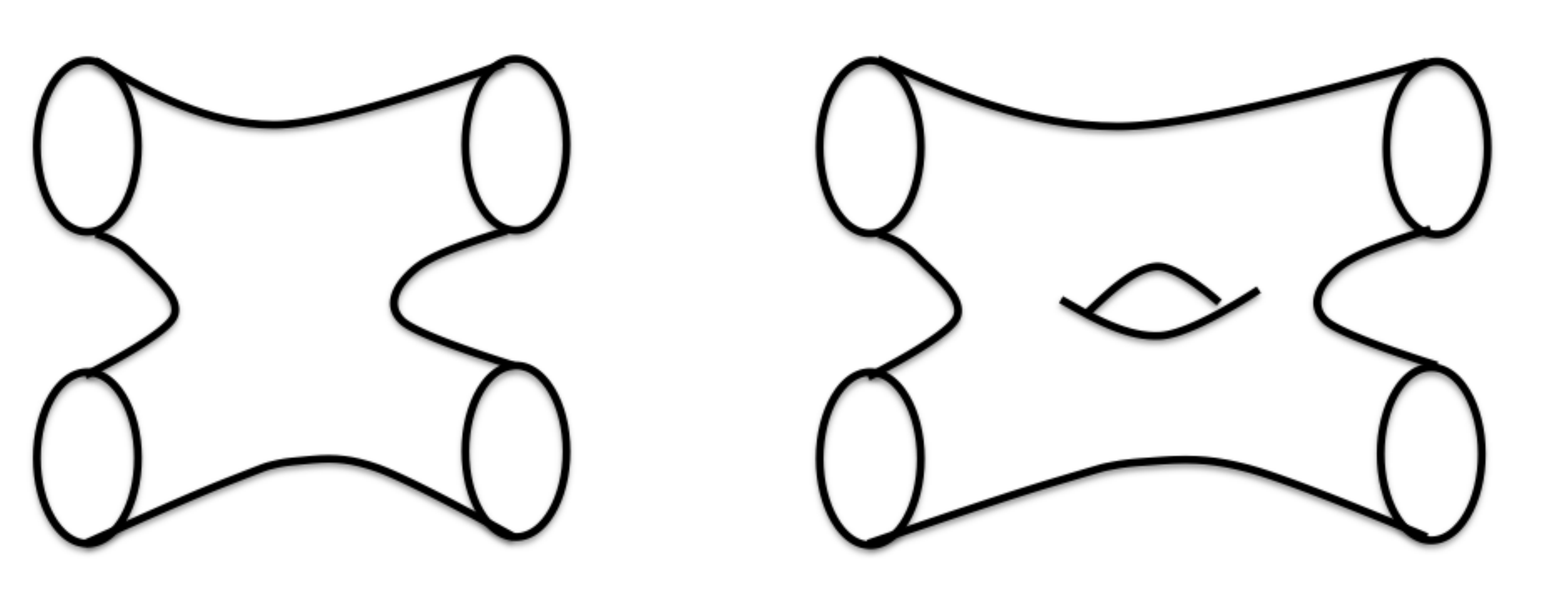}}
\caption{Scattering of two closed strings, tree-level (left) and one-loop (right).  Each loop involves a factor $g_s^2$. }\label{fig:string_scattering}
\end{center} 
\end{figure}

In the following I consider the type IIA and type IIB superstrings. There are two kinds of massless bosonic fields in these theories: NS-NS and R-R. 
The NS-NS fields include the graviton $G_{\mu\nu}$, the rank-two antisymmetric field $B_{\mu\nu}$ and the dilaton $\phi$. 
The expectation value of the dilaton is related to the string coupling constant $g_s$, which characterizes the strength of the quantum stringy effects, by $g_s=e^\phi$ 
(Fig.~\ref{fig:string_scattering}). 
The R-R fields are rank-$k$ antisymmetric tensor fields $C^{(k)}_{\mu_1\cdots\mu_k}$, where $k=1,3$ for IIA and $k=0,2,4$ for IIB. 
\begin{eqnarray}
\begin{array}{|c|c|c|}
\hline
 & type\ {\rm IIA}& type\ {\rm IIB}\\
\hline
{\rm NS-NS} & G_{\mu\nu}, B_{\mu\nu},\phi& G_{\mu\nu}, B_{\mu\nu},\phi \\
\hline
{\rm R-R} & C^{(1)}_{\mu}, C^{(3)}_{\mu\nu\rho} & C^{(0)}, C^{(2)}_{\mu\nu}, C^{(4)}_{\mu\nu\rho\sigma} \\
\hline
\end{array}
\nonumber
\end{eqnarray}

At low-energy, there is an effective description of the superstring theory in terms of the massless fields, 
which reproduces the scattering amplitudes calculated from string theory. This effective theory is the {\it supergravity}, 
which is a supersymmetric generalization of the Einstein gravity.  

\subsection{Black brane and D-brane}
Type IIA supergravity has a black hole solution with the R-R 1-form charge, which is called the black 0-brane. 
The black 0-brane preserves half of the supersymmetry (16 supercharges).  
In addition, there are $(1+p)$-dimensional supersymmetric black-hole-like objects, the black $p$-brane, 
which couples to the electric or magnetic charge of the R-R $(p+1)$-form.  Here $p$ is even in IIA and odd in IIB.  
The metric of the black $p$-brane is given by \cite{Gibbons:1987ps}  
\begin{eqnarray}
ds^2
&=&
\alpha'\Biggl\{
\frac{U^{(7-p)/2}}{g_{YM}\sqrt{d_p N}}
\left[
-\left(
1-\frac{U_0^{7-p}}{U^{7-p}}
\right)dt^2
+
dy_{\parallel}^2
\right]
\nonumber\\
& &
\qquad
+
\frac{g_{YM}\sqrt{d_p N}}{U^{(7-p)/2}\left(1-\frac{U_0^{7-p}}{U^{7-p}}\right)}dU^2
+
g_{YM}\sqrt{d_p N}U^{(p-3)/2}d\Omega_{8-p}^2
\Biggl\},
\label{near_extremal_metric} 
\end{eqnarray}
where we have used the Yang-Mills coupling $g_{YM}$ and the size of the gauge group $N$ in the corresponding super Yang-Mills theory 
for later convenience. 
A constant $\alpha'$ is the inverse of the string tension, 
$(t,y_\parallel)$ is the coordinate of the $(p+1)$-dimensions along which the brane is extended, 
$U$ and $\Omega$ are the radial and angular coordinate of the transverse directions, and  
$U_0$ is the place of the horizon, which is related to the Hawking temperature of the black brane by 
\begin{eqnarray}
T=\frac{(7-p)U_0^{(5-p)/2}}{4\pi \sqrt{d_p g_{YM}^2N}},  
\end{eqnarray}
where $d_p=2^{7-2p}\pi^{(9-3p)/2}\Gamma((7-p)/2)$. 
When $p=3$, the coefficient in front of $d\Omega_{8-p}^2$ becomes a constant independent of $U$, 
and hence the black brane geometry becomes a direct product of the five-dimensional spacetime 
described by $(t,y_\parallel,U)$ (AdS black hole) and the five sphere $S^5$. 
When $p\neq 3$, there is no simple direct product structure. 
The dilaton, or equivalently the string coupling constant, is given by 
\begin{eqnarray}
e^{\phi}=(2\pi)^{2-p}g_{YM}^2
\left(
\frac{d_p g_{YM}^2N}{U^{7-p}}
\right)^{\frac{3-p}{4}}. 
\label{string_coupling}
\end{eqnarray}
At $p=3$, it is independent of $U$. 

The black brane is a classical solution to the supergravity. Because the supergravity is the low-energy effective description of the superstring theory, 
the black brane must exist also in the superstring theory, at least in the parameter region where the supergravity gives a good approximation. 
Then, is it possible to describe the black brane directly in the string perturbation theory?  

\begin{figure}[htbp] 
\begin{center}
\scalebox{0.3}{\includegraphics{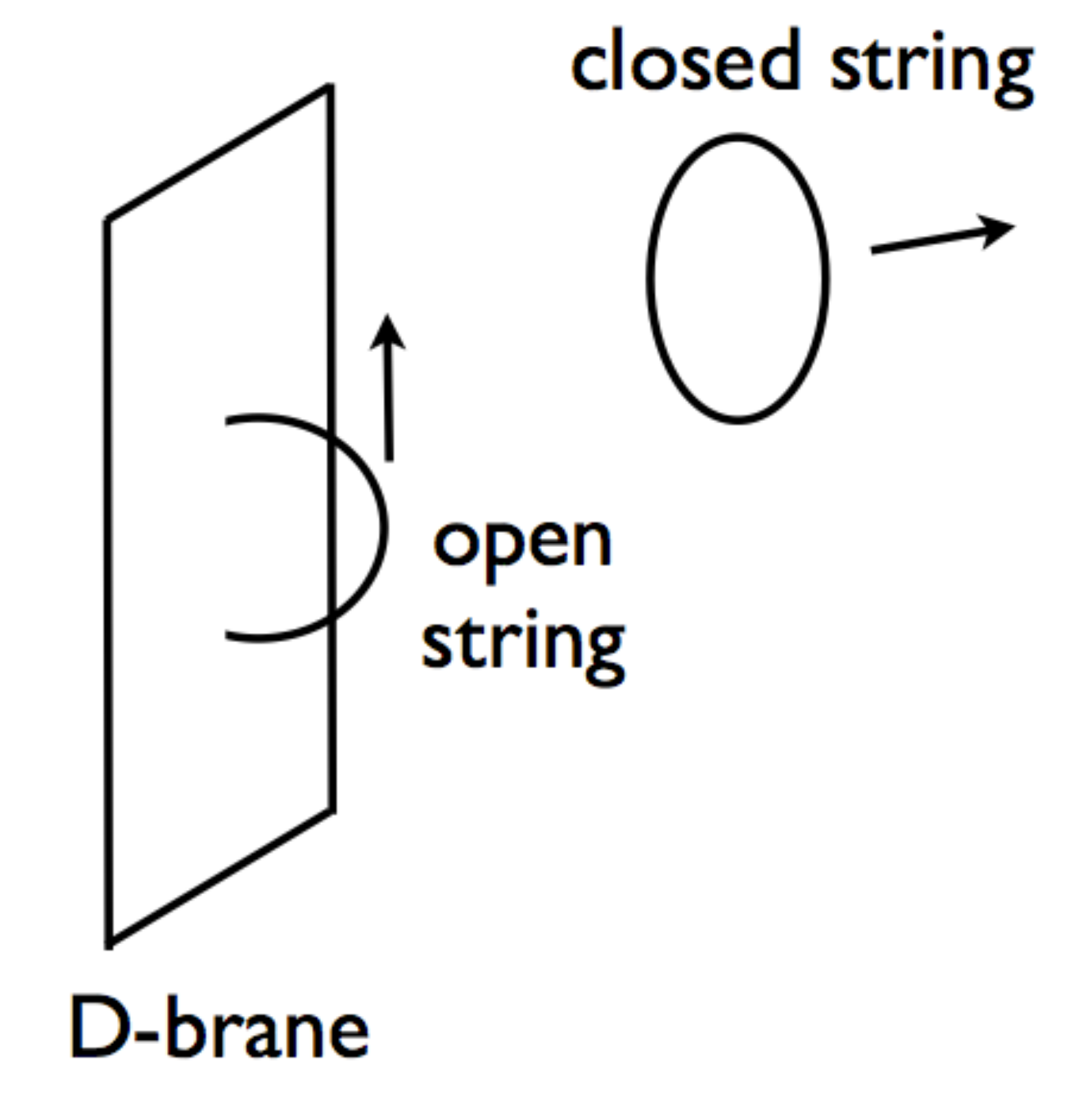}}
\caption{ 
An open string must be attached to D-branes. A closed string can propagate anywhere in the bulk.
}\label{fig:Dbrane_1}
\end{center} 
\end{figure}

Actually there is a counterpart, which is the D-brane (or `Dirichlet' brane) found by Polchinski and collaborators \cite{Dai:1989ua,Polchinski:1995mt}. 
The D$p$-brane is a membrane-like object extending to the time and $p$ spatial dimensions $x^0=t,x^1,\cdots,x^p$. 
(Again, $p$ is even in IIA and odd for IIB, in order for the supersymmetry to be preserved.) 
While the closed string can propagate anywhere in the bulk $(1+9)$-dimensional spacetime, 
the endpoints of the open string must be attached to the D-brane, 
or equivalently the Dirichlet (Neuman) boundary condition along $x^{p+1},\cdots,x^9$ ($x^0,x^1,\cdots,x^p$) 
is imposed for the endpoints (Fig.~\ref{fig:Dbrane_1}). 

\begin{figure}[htbp] 
\begin{center}
\scalebox{0.3}{\includegraphics{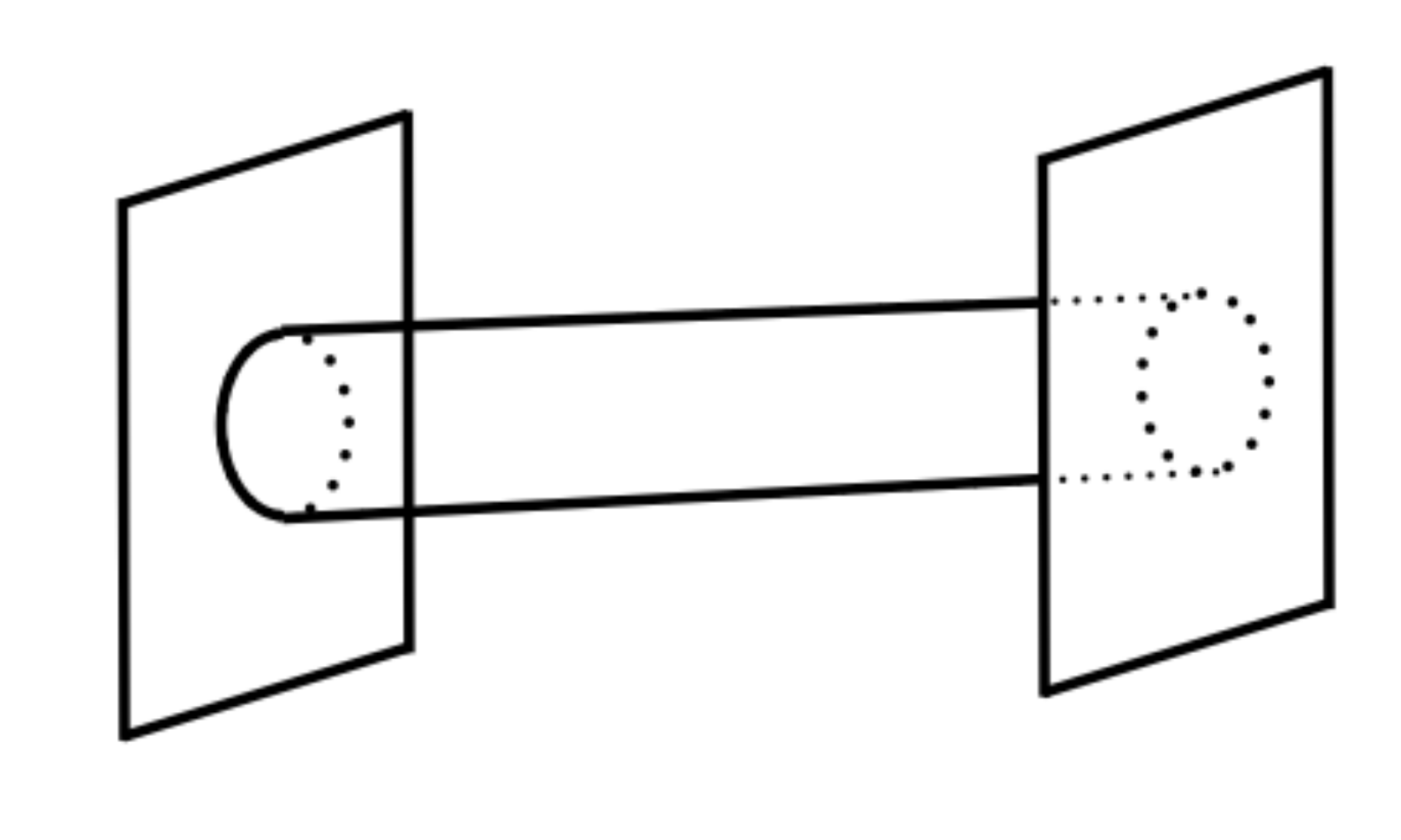}}
\caption{A cartoon picture of the open string-closed string duality. A loop of an open string stretching between two D-branes can also be interpreted as an exchange of a closed string.  }\label{fig:OpenClosedDuality}
\end{center} 
\end{figure}

In order to understand the physics of the black brane, one considers the emission or absorption of the graviton, which come from the closed string. 
On the other hand, the concept of the D-brane manifestly relies on the open string picture. Therefore, their relationship can be understood as the {\it open string-closed string duality}.  
This is intuitively very simple; as shown in Fig.~\ref{fig:OpenClosedDuality}, a loop of an open string stretching between two D-branes 
can also be interpreted as an exchange of a closed string. By using this simple picture, Polchinski showed that the D-brane has the same R-R charge 
as the black brane, and that it preserves the same supersymmetry, which is the evidence that the D-brane and the black brane are the same object.    


\subsection{D-brane and super Yang-Mills}
Let us consider the low-energy effective theory of the D-branes sitting in parallel and open strings stretching between them. 
(If they are not sitting in parallel, supersymmetry is broken in general.) 

First let us consider the single D-brane (Fig.~\ref{fig:Dbrane_1}). 
Because the open strings are trapped on the D-brane, the low-energy theory is a $(p+1)$-dimensional field theory. 
At sufficiently low energy, the oscillatory modes decouple and only the massless modes remain. 
They are the gauge field $A_\mu (\mu=0,1,\cdots,p)$, scalar fields $X_i (i=1,\cdots,9-p)$ and their superpartner $\psi_\alpha (\alpha=1,2,\cdots,16)$. 
Note that scalar fields describe the transverse fluctuation of the D-brane, which appears because the strings pull the D-brane. 

\begin{figure}[htbp] 
\begin{center}
\scalebox{0.3}{\includegraphics{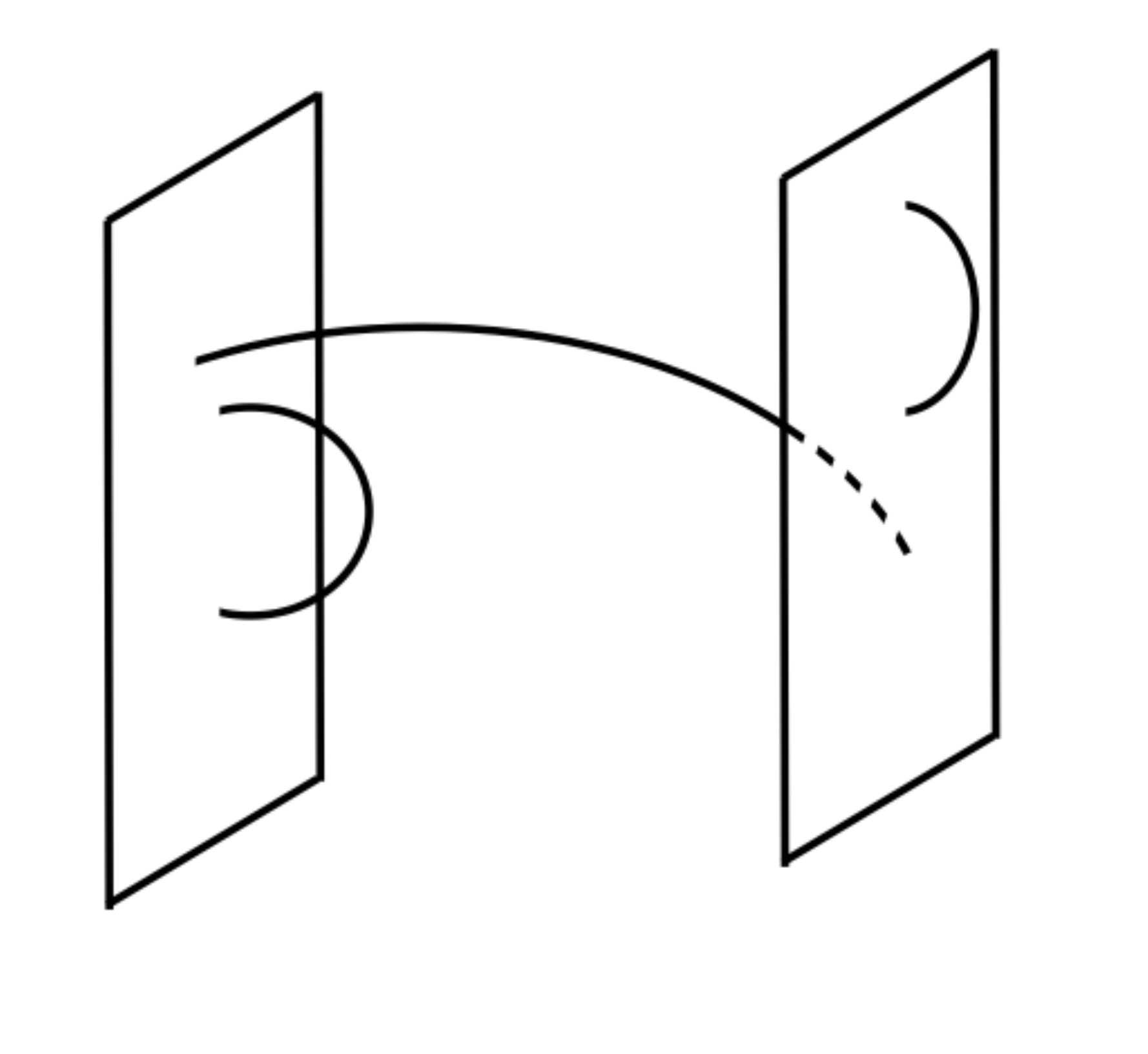}}
\caption{The system of two parallel D-branes and open strings.  Because string has the tension, the string connecting two different D-branes becomes massive.  }\label{fig:Dbrane_2}
\end{center} 
\end{figure}

Next let us consider the system of two D-branes (Fig.~\ref{fig:Dbrane_2}). Now there are two types of the open strings: 
the one having both ends on the same brane, and the one connecting two different branes. 
The former is massless, while the latter has the mass,  (the distance between D-branes)$\times$(the string tension).  
Such a situation is realized in the $U(2)$ Yang-Mills theory with adjoint matters, 
by identifying the $(i,j)$-component of the matrices with the strings connecting $i$-th and $j$-th D-branes.  
In fact, from the kinetic term of the $U(2)$ Yang-Mills 
\begin{eqnarray}
\frac{1}{2}\int d^{p+1}x\ Tr\left(D_\mu X_i\right)^2
\equiv
\frac{1}{2}\int d^{p+1}x\ Tr\left(\partial_\mu X_i - i[A_\mu,X_i]\right)^2, 
\end{eqnarray}
around the diagonal background $X_i=diag(x^{(1)}_i, x^{(2)}_i)$ the mass term 
$(x^{(1)}_i-x^{(2)}_i)^2\cdot |(A_\mu)_{12}|^2$ is obtained. Mass terms of the scalars and fermions appear  
from the interaction terms in the same manner. 
After a little bit more detailed argument, it turns out that the maximally supersymmetric $U(2)$ Yang-Mills in $(p+1)$-dimension\footnote{
In this proceeding we will consider the Euclidean theories, because we want to put them on computer. 
Therefore when we wrote `$(3+1)$-d' it should be interpreted as 4d Euclidean spacetime. The boundary condition is not necessarily thermal 
(i.e. anti-periodic b.c. for the fermion along the Euclidean time); we also consider the periodic boundary condition, which preserves the supersymmetry.  
}, 
\begin{eqnarray}
S=\frac{1}{g_{YM}^2}\int d^{p+1}x\ Tr\left\{
\frac{1}{4}F_{\mu\nu}^2
+
\frac{1}{2}\left(D_\mu X_i\right)^2
-
\frac{1}{4}
[X_i,X_j]^2
+
i\bar{\psi}\gamma^\mu D_\mu\psi
+
\bar{\psi}\gamma^i [X_i,\psi]
\right\}, 
\end{eqnarray}
gives the low-energy effective action. When there are $N$ D-branes, the $U(N)$ maximal super Yang-Mills appears. 
(Here `maximal' means the theory with 16 supercharges. If more supercharges exist, fields with spin larger than one 
necessarily appear, and it is difficult to construct a consistent local quantum field theory.) 

\subsection{The gauge/gravity duality}
Now we have two descriptions for the D$p$-branes: super Yang-Mills (open string picture) and the supergravity, or more precisely 
weakly coupled type IIA/IIB superstring theory, around the black $p$-brane (closed string picture). 
Can these descriptions become valid simultaneously? 

Because SYM takes into account only the open strings, one has to consider the limit where the closed strings decouple. 
Furthermore, because the SYM does not contain the oscillatory modes of the open strings, such modes must become heavy.  
Such a situation is realized when the D-branes are sitting very close to each other (the distance is much smaller than $\sqrt{\alpha'}$) 
and the physics near the horizon (closer than $\sqrt{\alpha'}$) is considered. 
In order for the weakly coupled closed string picture around the black brane background to make sense, 
the stringy corerction must be small. Therefore the curvature radius of the metric (\ref{near_extremal_metric}) 
must be large compared to $\sqrt{\alpha'}$ and the string coupling (\ref{string_coupling}) must be small.  
When $p=3$, these conditions are equivalent to $g_{YM}^2N\gg $ and $g_{YM}^2\ll 1$. 
In this parameter region, at least formally, SYM and weakly coupled IIA/IIB string theory picture become valid simultaneously. 
\footnote{Of course whether this condition is sufficient for the weakly coupled description is a highly nontrivial issue-- 
we are trying to apply a naive perturbative string theory picture to a very short distance!}  

Based on that observation, Maldacena conjectured these two theories are the same (Fig.~\ref{fig:AdS/CFT}). 
More precisely, the conjecture is as follows:


\begin{figure}[htbp] 
\begin{center}
\scalebox{0.25}{\includegraphics{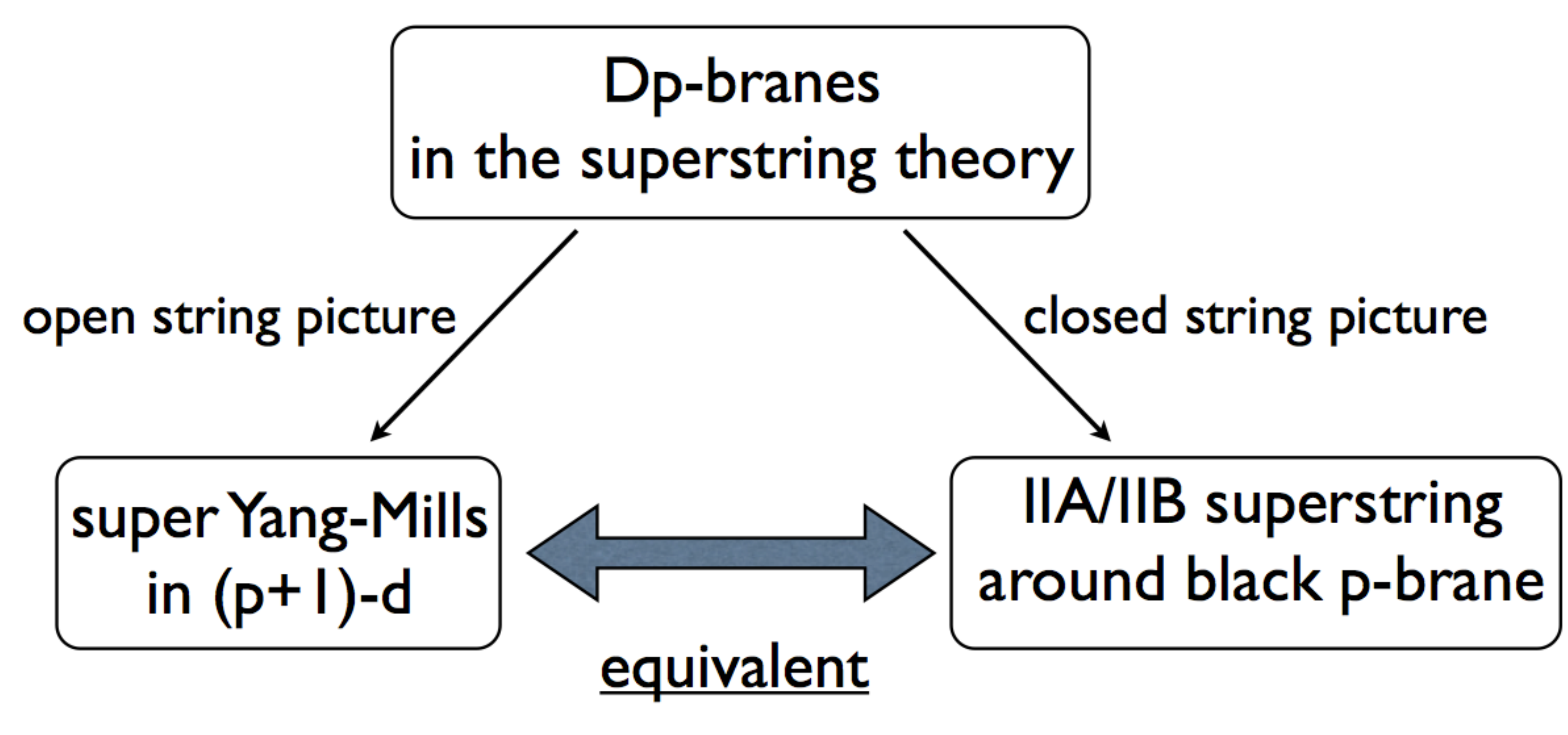}}
\caption{ 
The gauge/gravity duality. 
}\label{fig:AdS/CFT}
\end{center} \end{figure}

\begin{itemize}
\item
The large-$N$ and strong 't Hooft coupling limit of  SYM corresponds to the supergravity. 

\item
The large-$N$ and finite 't Hooft coupling corresponds to the classical string theory, i.e. $\alpha'> 0$ but $g_s=0$. 

\item
Finite-$N$, finite 't Hooft coupling corresponds to the full quantum string theory, $\alpha'> 0$ and $g_s> 0$. 

\end{itemize}
By now most string theorists believe the first two statements are true, while the last one is still controversial. 

Because the supergravity is just a classical theory and hence much easier than SYM, people often solve the supergravity to learn about SYM. 
However, once the stringy correction is taken into account, the gravity side suddenly becomes very difficult. 
If we can somehow study the gauge theory side, we can make various predictions for the string theory.  
Here lattice theorists can play important roles.

\section{Super Yang-Mills on computer}\label{sec:SYM}

\subsection{(No) fine tuning problem}
Even if a given lattice action becomes the target theory at tree level, it does not necessarily converge to the correct continuum limit at quantum level. 
In order to understand this point, let us recall why the Wilson's plaquette action works. 
The key is the {\it exact symmetries}; it has the exact gauge symmetry, discrete translation symmetry, 90 degree rotation symmetry, 
charge conjugation symmetry etc {\it at discretized level}.   
Therefore, radiative corrections must respect these symmetries, which assures the emergence of the correct continuum limit. 
If some of these exact symmetries are not kept, we might obtain a wrong continuum limit, which is not gauge invariant or translation invariant or whatever. 
The same problem is more familiar when one considers the fermion; with the Wilson fermion, for example, the mass-parameter fine tuning is needed 
because of the lack of the chiral symmetry on the lattice. 

The trouble with the lattice supersymmetry comes from the simple fact that the supersymmetry algebra contains the infinitesimal translation 
(the spacetime derivative), which is broken on lattice by construction. Therefore it is impossible to preserve the whole supersymmetry algebra. 
(For detailed arguments, see e.g. \cite{Kato:2008sp,Bergner:2009vg}.) 
Still, however, various theories allow fine-tuning free formulations: 
\begin{itemize}
\item
In 4d ${\cal N}=1$ pure SYM (i.e. a theory of gluons and gluinos, without matter fields), 
one can use the chiral symmetry to forbid SUSY breaking radiative corrections \cite{Kaplan:1983sk}. 

\item
In 2d extended SYM (i.e. the dimensional reductions of 4d ${\cal N}=1,2$ or $4$), 
one can preserve a part of the supersymmetry unbroken. This exact supersymmetry, combined with other global symmetries, 
assures the correct continuum limit. It was first argued at perturbative level \cite{Kaplan:2002wv}\cite{Cohen:2003xe,Sugino:2003yb,Catterall:2004np,D'Adda:2005zk} and 
then confirmed nonperturbatively by the Monte-Carlo simulation \cite{Kanamori:2008bk,Hanada:2009hq,Hanada:2010qg}. 

\item
For 3d and 4d SYM, matrix model techniques play important roles. In short, lattice-like structure can be embedded 
in big matrices, so that SYM theories are realized around special vacua of matrix models.  
In \cite{Maldacena:2002rb} and \cite{Hanada:2010kt,Takimi:2012zw}, it has been shown that the $(2+1)$-d and $(3+1)$-d $U(k)$ gauge theories 
can be obtained around the $k$-coincident fuzzy sphere background of $SU(Nk)$ theory in $(0+1)$d and $(1+1)$-d, after taking an appropriate large-$N$ limit.  
$(0+1)$-d and $(1+1)$-d theories can be regularized as we have seen before. 
Another formulation can be found in \cite{Ishii:2008ib}, which showed 
an equivalence between $(3+1)$-d SYM and a certain multiple fuzzy sphere background of the plane wave matrix model \cite{Berenstein:2002jq}. 
Among the formulations for $(3+1)$-d SYM, the one in \cite{Hanada:2010kt,Takimi:2012zw} has advantage that finite gauge group can be realized. 
On the other hand, although the method in \cite{Ishii:2008ib} is valid only for $U(\infty)$ gauge group, it preserve all the supersymmetry in a very elegant manner. 
(Note that, though four-dimensional lattice theories requires the fine tunings, however, the number of the fine tunings can be small 
and hence the fine tuning approach might be  practical. See \cite{Catterall:2011pd,Catterall:2012yq}.)

\item
For 1d theories, one does not have to worry about the radiative corrections so seriously\footnote{
One must not be too naive; see \cite{Giedt:2004vb}. 
}, thanks to the UV finiteness. 
It enables us to use a very efficient non-lattice method \cite{Hanada:2007ti}. 
 For a lattice approach, see \cite{Catterall:2007fp}. See also \cite{Wosiek:2002nm} for Hamiltonian formulation.  

\end{itemize}
Although several interesting theories like the supersymmetric QCD do not allow any fine-tuning free formulation yet, 
most theories important for the string theory can be put on computer without any fine tuning.

\subsection{(No) sign problem}
Another possible obstacle for the lattice SUSY simulation is the sign problem, that is, 
the determinant (or Pfaffian) of the Dirac operator is complex.  
However, rather surprisingly, it can simply be avoided by using the phase quenched ensemble 
in all examples so far studied.   

\begin{itemize}
\item
In 2d ${\cal N}=(2,2)$ SYM (i.e. the dimensional reduction of 4d ${\cal N}=1$ pure YM), 
there is no sign problem in the continuum limit. Although the complex phase appears as a lattice artifact \cite{Giedt:2003ve,Suzuki:2007jt}, 
one can take the correct continuum limit without the complex phase by using the phase quenched ensemble \cite{Hanada:2009hq,Hanada:2010qg,Catterall:2011aa}. 

\item
In 1d and 2d maximal SYM at finite temperature (i.e. antiperiodic boundary condition for the fermion along the temporal circle), 
the Pfaffian is almost real positive at high temperature region. 
Even at rather low temperature where a good agreement with the gravity dual is observed, the Pfaffian is still almost real and positive 
\cite{Anagnostopoulos:2007fw,Catterall:2008yz}\cite{Catterall:2010fx}.  

\item
With the periodic boundary condition, non-negligible phase fluctuation had been observed at $N\ge 3$ and/or sufficiently large volume.  
Still, however, the phase quench simulation gives consistent results with the gravity dual. 
When the average sign is not very close to zero, the validity of the phase quenching can also be confirmed numerically, without referring to the gravity dual; 
see \cite{Hanada:2011fq} for the detail.

\item
For a class of Wess-Zumino models, the world-line formalism and the worm algorithm can be used to avoid the sign problem
\cite{Baumgartner:2011cm}.

\end{itemize}

\subsection{$(0+1)$-d SYM and black hole thermodynamics}
Now let us see the simulation results and interpret them from string theory point of view.  
We start with the $(0+1)$-d maximal SYM, which is (conjectured to be) dual to type IIA string theory around the black zero-brane. 
Because the coupling constant has the dimension of $(mass)^3$, this theory is not conformal. However the gravity dual predicts 
a conformal-like behavior at large-$N$.  
For example, there are many massless modes whose correlators show a power-law decay, 
and this theory is deconfining (in the sense that the Polyakov loop has a nonzero expectation value) at any nonzero temperature. 
The Monte-Carlo study does not just reproduces such predictions but also provides us with further insights. 

\begin{itemize}
\item
The theory has a flat direction $[X_i,X_j]=0$. Because of the supersymmetry, it is not lifted by the quantum correction. 
However, if one takes the initial condition as $X_i=0$, a metastable state appears. That is, eigenvalues of $X_i$ fluctuate 
around zero for a while and then run to infinity. 
The metastable state becomes more and more stable as $N$ becomes larger. 
It is consistent with the stringy picture: 
because the scalars represent the D0-branes, 
the metastable state is the bound state of the D0-branes, which is regarded as the black zero-brane in the gravity picture. 
There is no instability in the black brane solution at classical level ($N=\infty$),  
and the instability appears through the quantum effect ($1/N$ correction). 

In order to take the expectation value, one has to take $N$ large enough, 
so that sufficiently many configurations are obtained before the metastable state collapses.  

\item
The thermal excitation energy $E$ in SYM corresponds to the ADM mass of the black hole.  
At large-$N$, on the gravity side, it is evaluated as 
\begin{eqnarray}
\frac{(\lambda^{-1/3}E)}{N^2}=7.41 (\lambda^{-1/3}T)^{2.8} + c (\lambda^{-1/3}T)^{4.6}+\cdots,  
\end{eqnarray}
where a constant $c$ has not yet determined analytically. 
Note that the 't Hooft coupling constant $\lambda=g_{YM}^2N$ has the dimension of $({\rm mass})^3$. 

Numerical study of this quantity had been performed in \cite{Anagnostopoulos:2007fw,Catterall:2008yz,Hanada:2008ez,Catterall:2009xn}. 
In particular,  in \cite{Hanada:2008ez} the next leading term had been studied. 
By using $\frac{(\lambda^{-1/3}E)}{N^2}=7.41 (\lambda^{-1/3}T)^{2.8} + c (\lambda^{-1/3}T)^{p}$ as the ansatz, 
we obtain $p=4.58(3)$ and $c=-5.55(7)$, which is consistent with the analytic prediction $p=4.6$. 
By a one-parameter fit with $p=4.6$, the overall coefficient is determined as $c=-5.58(1)$. 
This is a prediction from SYM, which should be reproduced from the string theory.  

\item
In the gauge/gravity duality, the supersymmetric Wilson loop which involves the scalar field can easily calculated on the gravity side \cite{Maldacena:1998im}. 
In $(0+1)$-d, one can consider the Wilson loop winding on the temporal circle (the Polyakov loop), which is given by 
$W=\frac{1}{N}Tr{\rm P}e^{\int dt(iA+X)}$, where $X$ is one of the scalars $X_1,\cdots,X_9$. 
The prediction from the gravity side is $\langle\log|W|\rangle\sim 1.89T^{-3/5} + C$, where $C$ is an unknown constant. 
The simulation result in the gauge theory is consistent with this prediction, by taking  $C=-4.95$ for $N=4$ and $-4.58$ for $N=6$ \cite{Hanada:2008gy}. 

\item
At large-$N$ and at strong coupling, 
correlation functions can be calculated by using the Gubser-Klebanov-Polyakov-Witten relation \cite{Gubser:1998bc} by solving the gravitational equations of motion.  
In \cite{Hanada:2009ne,Hanada:2011fq}, various two-point functions had been calculated numerically on the Gauge theory side, 
and perfect agreement with the dual gravity prediction \cite{Sekino:1999av} had been observed already at $N=2$.

\end{itemize}

\subsection{$(1+1)$-d SYM and black hole/black string transition}
Let us consider the $(1+1)$-dimensional $U(N)$ maximal SYM on spatial $S^1$ with circumference $L$. 
We impose the periodic boundary condition both for the bosonic and fermionic fields. 
This theory describes the system of $N$ D1-branes winding on the circle. 
By taking the T-duality along the circle, it can also describe the system of $N$ D0-branes.  
The compactification radius becomes $L'\equiv 2\pi\alpha'/L$. 
The positions of the D0-branes correspond to the eigenvalues of the Wilson line 
along the spatial circle, $W={\rm P}e^{i\int A_x dx}$. 
When the Wilson line phases clump to a point ($\langle W\rangle\neq 0$), they describe a bunch of D0-branes localized along the $S^1$, 
which is the black hole (black 0-brane) in the gravity picture. 
On the other hand, when the Wilson line phases are uniformly distributed on the unit circle ($\langle W\rangle = 0$), 
the D0-branes form the (uniform) `black string'. (Note that this black string is an object in type IIA string, which is different from the black 1-brane 
in type IIB.) One can also consider other objects like nonuniform black string and multiple black hole.  

\begin{figure}[htbp] 
\begin{center}
\scalebox{0.25}{\includegraphics{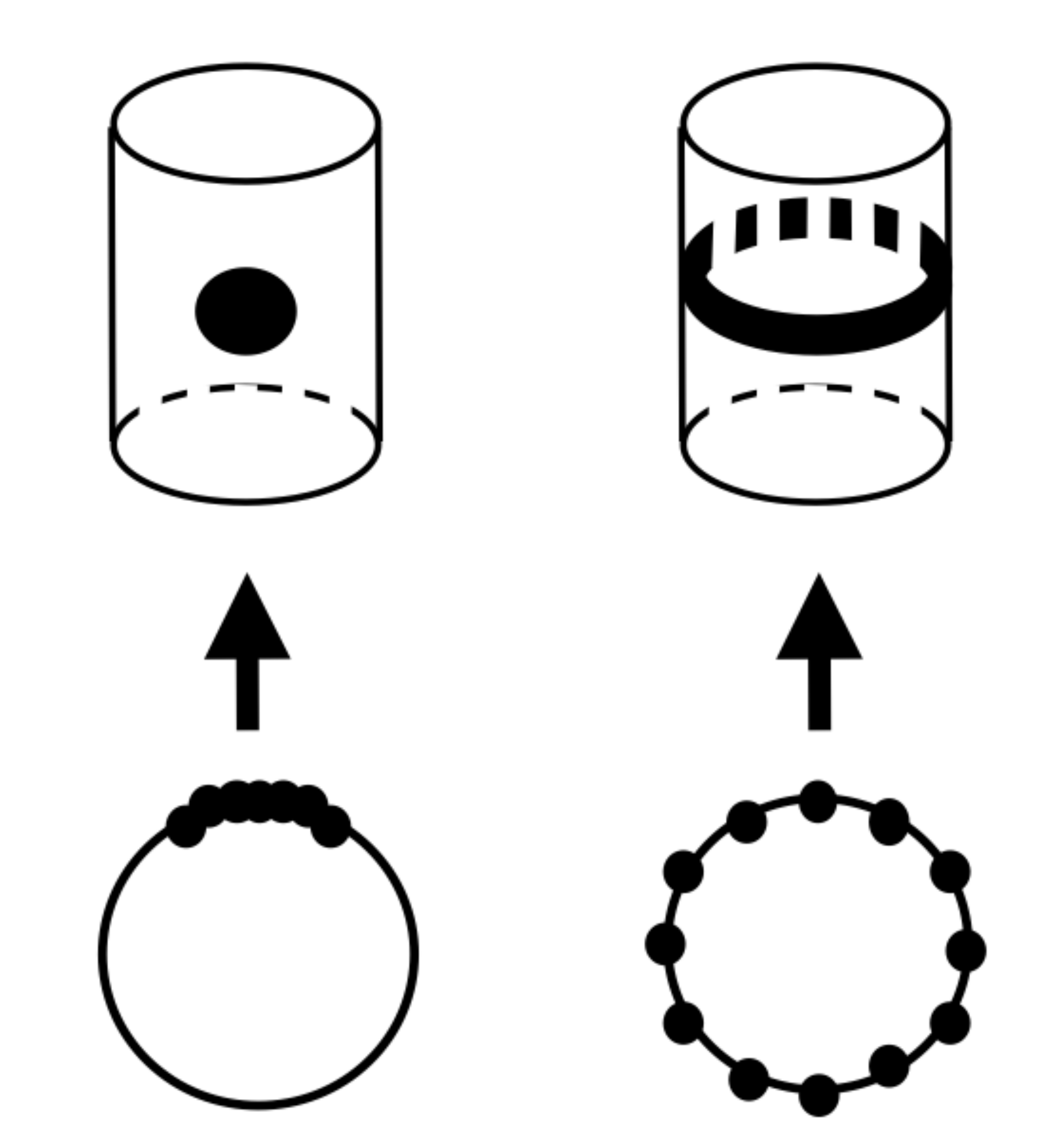}}
\caption{The center broken phase (left, bottom) and unbroken phase (right, bottom) 
	correspond to the black hole (left, top) and uniform black string (right, top). 
	The dots in the bottom figures represent the Wilson line phases, which take the values on the unit circle. 
	}\label{fig:BH/BS}
\end{center} 
\end{figure}

Let us fix the temperature and change the radius of the spatial circle. 
When the radius is sufficiently large, the black hole solution in the noncompact space provides us with a good approximation. 
As the radius becomes as small as the size of the black hole, the compactified direction gets completely filled by the black hole; 
then the black hole winds on the circle and turns into the black string. This is the black hole/black string transition \cite{Gregory:1993vy}. 
In the gauge theory, this transition can be identified with the breaking of the ${\mathbb Z}_N$ center symmetry $W\to e^{2\pi i/N}W$, 
because $W$ is an order parameter of the center symmetry (Fig.~\ref{fig:BH/BS}).    
By looking at the eigenvalue distribution of $W$ as a function of temperature $T$ and compactification radius $L$, one can study the stringy effect on the transition.   

At low temperature, the phase transition can be studied from the gravity side, 
by comparing the free energy of the black zero-brane and black string solutions to type IIA supergravity \cite{Aharony:2004ig}. 
(At small volume, strings winding on the compact circle become light, so that the IIB supergravity approximation is not good. 
Then by taking the T-dual, the winding modes are mapped to the KK momentum, and  description with IIA supergravity becomes valid. 
It can be shown that the phase transition takes place where type IIA supergravity, rather than type IIB supergravity, gives a good approximation.)
It turns out the black hole (center broken phase) is preferred at small $L$, and 
there is a first order transition to the uniform black string (center unbroken phase).  

At high temperature, because all the KK modes along the temporal direction (and hence all the fermionic degrees of freedom) decouple, 
the system reduces to the bosonic matrix quantum mechanics (bMQM). 
Note that the spatial $S^1$ in the two-dimensional theory corresponds to the temporal $S^1$ in the bMQM; 
hence the deconfinement transition of the latter can be regarded as a remnant of the black hole/black string transition.  
See \cite{Aharony:2004ig,Kawahara:2007fn} for detail studies of bMQM from this point of view.  

In \cite{Catterall:2010fx}, the phase structure had been studied by using the twisted SUSY formulation.  
Although larger scale simulation with larger $N$ and finer lattice is needed for rigid conclusion\footnote{
Because $N$ is not large enough to stabilize the flat direction, a small scalar mass is introduced. 
}, 
the result seems to interpolate known results in high and low temperature regions, and looks rather convincing. 
It is very important to pursue this direction further by using several lattice formulations. 

\subsection{Toward the simulation of $(3+1)$-d maximal SYM}
Although a simulation of the four-dimensional maximal SYM is very expensive, there is steady progress. 
\begin{itemize}
\item
In \cite{Honda:2010nx,Honda:2011qk}, a numerical simulation had been performed by using the large-$N$ equivalence 
between a special vacuum of a $(0+1)$-dimensional matrix model and the 4d SYM \cite{Ishii:2008ib}. 
They had evaluated certain Wilson loops \cite{Honda:2010nx} and correlation functions \cite{Honda:2011qk}, 
and confirmed the prediction of the AdS/CFT duality. 

\item
As argued in \cite{Catterall:2011aa}, although all known lattice formulations have a fine-tuning problem, 
the lattice simulation may still be manageable.  
After the conference an interesting paper \cite{Catterall:2012yq} appeared, in which a lattice simulation had been performed with this philosophy. 
%
As a first guess, the simulation has been performed without any parameter fine tunings.   
At all bare couplings they studied, their results look consistent with the theoretical predictions.   
The next important step should be to see the restoration of the supersymmetry, with a possible parameter fine tuning.

\item
Four-dimensional SYM can also be regularized by combining the two-dimensional lattice and fuzzy sphere  
\cite{Hanada:2010kt,Takimi:2012zw}. Because the same technique is applicable for the 3d SYM \cite{Maldacena:2002rb}, 
which is numerically much cheaper, it should be easier to test the validity of this method at nonperturbative level 
by a simulation of 3d SYM.

\item
It has been argued that physics of the supersymmetric sector in the strong coupling region can be captured by a simple matrix model \cite{Berenstein:2005aa}. 
For numerical work along this direction, see e.g. \cite{Berenstein:2008jn}.

\end{itemize}

\section{ABJM and membrane mini-revolution}\label{sec:ABJM}
The gauge/gravity duality is expected to hold also in M-theory, whose precise definition is yet to be known. 
In particular, a stack of multiple M2-branes (i.e. the usual supermembranes in eleven dimension \cite{Bergshoeff:1987cm})  
should be described by a three-dimensional supersymmetric conformal field theory, 
whose dual gravitational description is the M-theory on the $AdS_4\times S^7$ background. 
Such a theory had been found in 2008 by Aharony, Bergman, Jafferis and Maldacena (ABJM) \cite{Aharony:2008ug}. 
It is a $U(N)\times U(N)$ Chern-Simons theory with bifundamental matter fields, which is rather complicated. 
Its coupling constant, or equivalently the Chern-Simons level, is quantized to be integer $k$. 
M-theory on $AdS_4\times S^7/{\mathbb Z}_k$ gives a good description when $N\gg k^5$, 
while type IIA superstring theory on $AdS_4\times S^7/{\mathbb C}P_3$ is good when $k\ll N\ll k^5$.  
The dual M-theory geometry predicts a peculiar scaling of the free energy, $F\sim \sqrt{k}N^{3/2}$. 

It looks hopelessly difficult to study this system on lattice. However, rather unexpectedly, 
Monte Carlo approach turned out to be very effective for studying this system. 

The key is the localization method \cite{Pestun:2007rz,Kapustin:2009kz}, 
which maps the ABJM theory to a much simpler matrix model.  
Let us consider the partition function, 
\begin{eqnarray}
Z=\int [d\phi] e^{-S[\phi]}. 
\end{eqnarray}
Suppose the action and the path-integral measure are invariant under a fermionic symmetry generated by $Q$. 
We consider a one-parameter deformation, $S_t\equiv S+tQV$, where $V$ is an arbitrary fermionic function and $0\ge t$, and consider 
$Z_t\equiv \int [d\phi] e^{-S_t}$. Then 
\begin{eqnarray}
\frac{d}{dt}Z_t
=
 \int [d\phi] QV e^{-S_t}
 =
 \int [d\phi]Q(Ve^{-tQV})e^{-S}
 =
 Z\times\langle Q(Ve^{-tQV})\rangle
 =
 0,  
\end{eqnarray}
and hence $Z_t$ is actually $t$-independent. In the limit $t\to\infty$, $tQV$ gives an infinitely steep potential, 
the path-integral `localizes' to the classical solution to this potential, and one-loop calculation around the classical solution 
provides us with the exact answer. With a clever choice of $V$, the classical solution becomes very simple, sometimes as simple as one matrix, 
and then the partition function $Z$ is the same as that of the matrix model. 
In 4d maximal SYM,  for example, the resultant matrix model is simply Gaussian \cite{Pestun:2007rz}. 
The same argument can be repeated for the expectation values of the $Q$-invariant operators; for example, a supersymmetric Wilson loop 
in 4d maximal SYM can be calculated by using the Gaussian matrix model. 

The localization method had been applied to the ABJM theory in \cite{Kapustin:2009kz}. The partition function can be written as 
\begin{eqnarray} 
Z^\mathrm{ABJM}
\,=\,
\frac{1}{(N!)^2}\int\frac{d^N\mu}{(2\pi)^N}\frac{d^N\nu}{(2\pi)^N}
\frac{\prod_{i<j}\Bigl(2\sinh\frac{\mu_i-\mu_j}{2}\Bigr)^2 \Bigl(2\sinh\frac{\nu_i-\nu_j}{2}\Bigr)^2}
{\prod_{i,j} \Bigl(2\cosh\frac{\mu_i-\nu_i}{2}\Bigr)^2}
\exp\left[\frac{ik}{4\pi}\sum_{i=1}^N (\mu_i^2-\nu_i^2)\right]. 
\nonumber\\ 
\end{eqnarray}

Analytic studies of this theory confirmed $F\sim \sqrt{k}N^{3/2}$ including the prefactor \cite{Drukker:2010nc}, 
and important result on the $1/N$ correction had been obtained \cite{Fuji:2011km}. 
On the other hand by using the standard Monte Carlo technique this integral can be studied much more easily \cite{Hanada:2012si}; 
because the simulation is so cheap, we can obtain quite accurate numerical values.    
As a result, by combining the analytic and numerical techniques 
the analytic expression to all order in $1/N$-expansion, which corresponds the string perturbation theory,  had been obtained \cite{Fuji:2011km,Hanada:2012si}.  
After the lattice conference, analytic techniques to study this integral has been developed further; exact values are obtained for various $N$ and $k$ 
\cite{Hatsuda:2012hm}, and by subtracting the perturbative contribution the instanton effect has been discussed \cite{Hatsuda:2012dt}. 
Now the gauge theory side is under perfect control, and the quantum correction on the gravity side is going on,  
in order to test the AdS/CFT duality at fully quantum level. 
It is interesting to apply similar numerical techniques to other important theories.

\section{Other topics}\label{sec:others}
There are many other interesting developments in the numerical simulations. Here is an incomplete list\footnote{
Let me know if you find something missing here.
}:

\begin{itemize}

\item
4d ${\cal N}=1$ minimal super Yang-Mills has been actively studied \cite{Giedt:2008xm}.  
Probably the first target is the spontaneous breakdown of the ${\mathbb Z}_{2N}$ symmetry (here $N$ is the size of the gauge group $SU(N)$) 
due to the gaugino condensation. 
(Interesting arguments on SYM multiplet on the lattice \cite{Suzuki:2012gi} appeared after the lattice conference.) 

\item
Spontaneous SUSY breaking is also a very important subject. 
Hori and Tong \cite{Hori:2006dk} conjectured that in the two-dimensional ${\cal N}=(2,2)$ SYM, which is obtained from 4d ${\cal N}=1$ minimal SYM 
through the dimensional reduction, the supersymmetry is spontaneously broken. Although numerical work by Kanamori et al \cite{Kanamori:2007yx} 
is consistent with {\it unbroken} SUSY, it is not clear whether they observe the same vacuum Hori and Tong considered\footnote{
This theory has several (perhaps metastable) vacua \cite{Hanada:2009hq}; in the one Hori and Tong considered, eigenvalues of the scalar field 
are bounded around zero.   
}, and further study is needed. 

\item
The Wess-Zumino models on lattice are also actively studied. See 
\cite{Bergner:2007pu,Kadoh:2009sp} for the restoration of supersymmetry in the continuum limit, 
\cite{Kastner:2008zc} for the RG flow, 
\cite{Kawai:2010yj,Kamata:2011fr} for a use of the Nicolai map and the relation to a certain superconformal minimal model,  
\cite{Schierenberg:2012pb} for an improvement of the lattice actions of the matrix quantum mechanics,  
and \cite{Wozar:2011gu,Baumgartner:2011cm} for the spontaneous SUSY breaking.

\item
In \cite{Kim:2011cr} the ground state structure of the IKKT matrix model \cite{Ishibashi:1996xs}, which is the dimensional reduction of 
the 4d ${\cal N}=4$ SYM to zero dimension, has been studied. 
In this model, the distribution of the eigenvalues may be regarded as the spacetime; there are ten matrices, and if the eigenvalue distribution is four-dimensional 
it might be regarded as the emergence of our four-dimensional spacetime. 
In a sharp contrast to previous studies, they considered the Lorentzian version directly, and found that one time direction and three expanding 
spacial dimensions appear. 

Emergence of the spacetime from the point of view of the noncommutative geometry is also actively studied. 
(See e.g. \cite{Steinacker:2010rh,Chatzistavrakidis:2010tq} for recent progress. ) 
It would be interesting to apply numerical techniques to such scenarios.

\end{itemize}

\section*{Acknowledgement}
I would like to thank M.~Honda, I.~Kanamori, J.~Nishimura and S.~Shiba for useful comments on the draft.


\begin{thebibliography}{99}
  

\bibitem{Banks:1996vh}
T.~Banks, W.~Fischler, S.~H.~Shenker and L.~Susskind,
Phys.\ Rev.\ D {\bf 55}, 5112 (1997).     

\bibitem{Ishibashi:1996xs}
N.~Ishibashi, H.~Kawai, Y.~Kitazawa and A.~Tsuchiya,
Nucl.\ Phys.\ B {\bf 498}, 467 (1997).   

\bibitem{MatrixString}
  L.~Motl,
  arXiv:hep-th/9701025;   
  %
 R.~Dijkgraaf, E.~P.~Verlinde and H.~L.~Verlinde,
  Nucl.\ Phys.\  B {\bf 500}, 43 (1997).   


\bibitem{Maldacena:1997re}
  J.~M.~Maldacena, 
Adv.\ Theor.\ Math.\ Phys.\  {\bf 2}, 231 (1998).  

\bibitem{Itzhaki:1998dd}
  N.~Itzhaki, J.~M.~Maldacena, J.~Sonnenschein and S.~Yankielowicz,
Phys.\ Rev.\ D {\bf 58}, 046004 (1998).


\bibitem{Catterall:2009it} 
  S.~Catterall, D.~B.~Kaplan and M.~Unsal,
  Phys.\ Rept.\  {\bf 484}, 71 (2009). 

\bibitem{Nishimura:2012xs} 
  J.~Nishimura,
  arXiv:1205.6870 [hep-lat].
  
  
\bibitem{Catterall:2010nd} 
  S.~Catterall,
  arXiv:1005.5346 [hep-lat].

\bibitem{Gibbons:1987ps} 
  G.~W.~Gibbons and K.~-i.~Maeda,
  Nucl.\ Phys.\ B {\bf 298}, 741 (1988); 
  G.~T.~Horowitz and A.~Strominger,
  Nucl.\ Phys.\ B {\bf 360}, 197 (1991).
  

  
\bibitem{Dai:1989ua} 
  J.~Dai, R.~G.~Leigh and J.~Polchinski,
  Mod.\ Phys.\ Lett.\ A {\bf 4}, 2073 (1989).
  
\bibitem{Polchinski:1995mt} 
  J.~Polchinski,
  Phys.\ Rev.\ Lett.\  {\bf 75}, 4724 (1995). 

\bibitem{Kato:2008sp} 
  M.~Kato, M.~Sakamoto and H.~So,
  JHEP {\bf 0805}, 057 (2008). 
  
\bibitem{Bergner:2009vg} 
  G.~Bergner, JHEP {\bf 1001}, 024 (2010).
  

  
\bibitem{Kaplan:1983sk} 
  D.~B.~Kaplan,
  Phys.\ Lett.\ B {\bf 136}, 162 (1984); 
  %
  G.~Curci and G.~Veneziano,
  Nucl.\ Phys.\ B {\bf 292}, 555 (1987).
  
\bibitem{Kaplan:2002wv}
  D.~B.~Kaplan, E.~Katz and M.~\"{U}nsal,
  JHEP {\bf 0305}, 037 (2003).   
 
 \bibitem{Cohen:2003xe}
  A.~G.~Cohen, D.~B.~Kaplan, E.~Katz and M.~\"{U}nsal,
  JHEP {\bf 0308}, 024 (2003);   
  %
  JHEP {\bf 0312}, 031 (2003).   
%
  D.~B.~Kaplan and M.~\"{U}nsal,
  JHEP {\bf 0509}, 042 (2005).  
%
%
\bibitem{Sugino:2003yb}
  F.~Sugino,
  JHEP {\bf 0401}, 015 (2004);   
  %
  JHEP {\bf 0403}, 067 (2004);   
%
  JHEP {\bf 0501}, 016 (2005).   
  %
\bibitem{Catterall:2004np}
  S.~Catterall,
  JHEP {\bf 0411}, 006 (2004).   
%
  S.~Catterall,
  JHEP {\bf 0506}, 027 (2005). 
%
\bibitem{D'Adda:2005zk}
  A.~D'Adda, I.~Kanamori, N.~Kawamoto and K.~Nagata,
  Phys.\ Lett.\  B {\bf 633}, 645 (2006).   


 
\bibitem{Kanamori:2008bk} 
  I.~Kanamori and H.~Suzuki,
  Nucl.\ Phys.\ B {\bf 811}, 420 (2009). 
  
  
\bibitem{Hanada:2009hq}
  M.~Hanada and I.~Kanamori,
  Phys.\ Rev.\  D {\bf 80}, 065014 (2009). 
  
  
\bibitem{Hanada:2010qg}
  M.~Hanada and I.~Kanamori,
  JHEP {\bf 1101}, 058 (2011). 

\bibitem{Maldacena:2002rb} 
  J.~M.~Maldacena, M.~M.~Sheikh-Jabbari and M.~Van Raamsdonk,
  JHEP {\bf 0301}, 038 (2003). 
 
\bibitem{Hanada:2010kt} 
  M.~Hanada, S.~Matsuura and F.~Sugino,
  Prog.\ Theor.\ Phys.\  {\bf 126}, 597 (2011);  
%
  M.~Hanada,
  JHEP {\bf 1011}, 112 (2010). 

\bibitem{Takimi:2012zw} 
  T.~Takimi,
  JHEP {\bf 1208}, 069 (2012). 
  
\bibitem{Ishii:2008ib}
  T.~Ishii, G.~Ishiki, S.~Shimasaki and A.~Tsuchiya,
  Phys.\ Rev.\  D {\bf 78}, 106001 (2008). 
  
\bibitem{Berenstein:2002jq} 
  D.~E.~Berenstein, J.~M.~Maldacena and H.~S.~Nastase,
  JHEP {\bf 0204}, 013 (2002)
  [hep-th/0202021].
  
\bibitem{Catterall:2011pd} 
  S.~Catterall, E.~Dzienkowski, J.~Giedt, A.~Joseph and R.~Wells,
  JHEP {\bf 1104}, 074 (2011). 

\bibitem{Catterall:2012yq} 
  S.~Catterall, P.~H.~Damgaard, T.~Degrand, R.~Galvez and D.~Mehta,
  arXiv:1209.5285 [hep-lat].
   
  
\bibitem{Giedt:2004vb} 
  J.~Giedt, R.~Koniuk, E.~Poppitz and T.~Yavin,
  JHEP {\bf 0412}, 033 (2004). 
  
\bibitem{Hanada:2007ti} 
  M.~Hanada, J.~Nishimura and S.~Takeuchi,
  Phys.\ Rev.\ Lett.\  {\bf 99}, 161602 (2007). 
  

\bibitem{Catterall:2007fp} 
  S.~Catterall and T.~Wiseman,
  JHEP {\bf 0712}, 104 (2007).  
%

\bibitem{Wosiek:2002nm} 
  J.~Wosiek,
  Nucl.\ Phys.\ B {\bf 644}, 85 (2002)
  [hep-th/0203116].


\bibitem{Giedt:2003ve} 
  J.~Giedt,
  Nucl.\ Phys.\ B {\bf 668}, 138 (2003). 
  
\bibitem{Suzuki:2007jt} 
  H.~Suzuki,
  JHEP {\bf 0709}, 052 (2007). 
  
\bibitem{Catterall:2011aa} 
  S.~Catterall, R.~Galvez, A.~Joseph and D.~Mehta,
  JHEP {\bf 1201}, 108 (2012). 
  
\bibitem{Anagnostopoulos:2007fw} 
  K.~N.~Anagnostopoulos, M.~Hanada, J.~Nishimura and S.~Takeuchi,
  Phys.\ Rev.\ Lett.\  {\bf 100}, 021601 (2008). 

\bibitem{Catterall:2008yz}
  S.~Catterall and T.~Wiseman,
  Phys.\ Rev.\  D {\bf 78}, 041502 (2008). 


\bibitem{Catterall:2010fx} 
  S.~Catterall, A.~Joseph and T.~Wiseman,
  JHEP {\bf 1012}, 022 (2010). 
  
\bibitem{Hanada:2011fq} 
  M.~Hanada, J.~Nishimura, Y.~Sekino and T.~Yoneya,
  JHEP {\bf 1112}, 020 (2011). 
 
\bibitem{Baumgartner:2011cm} 
  D.~Baumgartner and U.~Wenger,
  arXiv:1104.0213 [hep-lat]; 
%
  D.~Baumgartner, K.~Steinhauer and U.~Wenger,
  PoS LATTICE {\bf 2011}, 253 (2011); 
  U.~Wenger,
  Phys.\ Rev.\ D {\bf 80}, 071503 (2009). 
 
\bibitem{Hanada:2008ez} 
  M.~Hanada, Y.~Hyakutake, J.~Nishimura and S.~Takeuchi,
  Phys.\ Rev.\ Lett.\  {\bf 102}, 191602 (2009). 
  
\bibitem{Catterall:2009xn} 
  S.~Catterall and T.~Wiseman,
  JHEP {\bf 1004}, 077 (2010). 
  
\bibitem{Maldacena:1998im} 
  J.~M.~Maldacena,
  Phys.\ Rev.\ Lett.\  {\bf 80}, 4859 (1998); 
%
  S.~-J.~Rey and J.~-T.~Yee,
  Eur.\ Phys.\ J.\ C {\bf 22}, 379 (2001). 
  
\bibitem{Hanada:2008gy} 
  M.~Hanada, A.~Miwa, J.~Nishimura and S.~Takeuchi,
  Phys.\ Rev.\ Lett.\  {\bf 102}, 181602 (2009). 
   
\bibitem{Gubser:1998bc} 
  S.~S.~Gubser, I.~R.~Klebanov and A.~M.~Polyakov,
  Phys.\ Lett.\ B {\bf 428}, 105 (1998); 
  %
  E.~Witten,
  Adv.\ Theor.\ Math.\ Phys.\  {\bf 2}, 253 (1998). 
 
\bibitem{Hanada:2009ne} 
  M.~Hanada, J.~Nishimura, Y.~Sekino and T.~Yoneya,
  Phys.\ Rev.\ Lett.\  {\bf 104}, 151601 (2010). 


\bibitem{Sekino:1999av} 
  Y.~Sekino and T.~Yoneya,
  Nucl.\ Phys.\ B {\bf 570}, 174 (2000). 
  

\bibitem{Gregory:1993vy}
  R.~Gregory and R.~Laflamme,
  Phys.\ Rev.\ Lett.\  {\bf 70}, 2837 (1993). 
  
 
\bibitem{Aharony:2004ig}
  O.~Aharony, J.~Marsano, S.~Minwalla and T.~Wiseman,
  Class.\ Quant.\ Grav.\  {\bf 21}, 5169 (2004).  

\bibitem{Kawahara:2007fn}
  N.~Kawahara, J.~Nishimura and S.~Takeuchi,
  JHEP {\bf 0710}, 097 (2007); 
%
  G.~Mandal, M.~Mahato and T.~Morita,
  JHEP {\bf 1002}, 034 (2010). 
  
 
\bibitem{Honda:2010nx}
  M.~Honda, G.~Ishiki, S.~-W.~Kim, J.~Nishimura and A.~Tsuchiya,
  PoS LATTICE {\bf 2010} (2010) 253;  
   %
\bibitem{Honda:2011qk} 
  M.~Honda, G.~Ishiki, J.~Nishimura and A.~Tsuchiya,
  PoS LATTICE {\bf 2011}, 244 (2011). 



\bibitem{Berenstein:2005aa} 
  D.~Berenstein,
  JHEP {\bf 0601}, 125 (2006). 



\bibitem{Berenstein:2008jn} 
  D.~Berenstein, R.~Cotta and R.~Leonardi,
  Phys.\ Rev.\ D {\bf 78}, 025008 (2008). 


\bibitem{Bergshoeff:1987cm} 
  E.~Bergshoeff, E.~Sezgin and P.~K.~Townsend,
  Phys.\ Lett.\ B {\bf 189}, 75 (1987).
  
\bibitem{Aharony:2008ug} 
  O.~Aharony, O.~Bergman, D.~L.~Jafferis and J.~Maldacena,
  JHEP {\bf 0810}, 091 (2008). 
  
  
\bibitem{Pestun:2007rz} 
  V.~Pestun,
  Commun.\ Math.\ Phys.\  {\bf 313}, 71 (2012). 
  
\bibitem{Kapustin:2009kz} 
  A.~Kapustin, B.~Willett and I.~Yaakov,
  JHEP {\bf 1003}, 089 (2010). 
 
  

\bibitem{Drukker:2010nc} 
  N.~Drukker, M.~Marino and P.~Putrov,
  Commun.\ Math.\ Phys.\  {\bf 306}, 511 (2011); 
  JHEP {\bf 1111}, 141 (2011). 


\bibitem{Fuji:2011km} 
  H.~Fuji, S.~Hirano and S.~Moriyama,
  JHEP {\bf 1108}, 001 (2011); 
%
  M.~Marino and P.~Putrov,
  J.\ Stat.\ Mech.\  {\bf 1203}, P03001 (2012). 

  
  
\bibitem{Hanada:2012si} 
  M.~Hanada, M.~Honda, Y.~Honma, J.~Nishimura, S.~Shiba and Y.~Yoshida,
  JHEP {\bf 1205}, 121 (2012). 
  
\bibitem{Hatsuda:2012hm} 
  Y.~Hatsuda, S.~Moriyama and K.~Okuyama,
  JHEP {\bf 1210}, 020 (2012); 
%
  P.~Putrov and M.~Yamazaki,
  Mod.\ Phys.\ Lett.\ A {\bf 27}, 1250200 (2012). 
  
\bibitem{Hatsuda:2012dt} 
  Y.~Hatsuda, S.~Moriyama and K.~Okuyama,
  arXiv:1211.1251 [hep-th].
  
\bibitem{Giedt:2008xm}
  J.~Giedt, R.~Brower, S.~Catterall, G.~T.~Fleming and P.~Vranas,
  Phys.\ Rev.\  D {\bf 79}, 025015 (2009); 
  M.~G.~Endres,
  Phys.\ Rev.\  D {\bf 79}, 094503 (2009);  
  K.~Demmouche, F.~Farchioni, A.~Ferling, I.~Montvay, G.~Munster, E.~E.~Scholz, J.~Wuilloud,
  Eur.\ Phys.\ J.\  {\bf C69}, 147-157 (2010); 
%
  G.~Bergner, I.~Montvay, G.~Munster, D.~Sandbrink and U.~D.~Ozugurel,
  arXiv:1210.7767 [hep-lat]; 
  S.~W.~Kim {\it et al.}  [JLQCD Collaboration],
  PoS LATTICE {\bf 2011}, 069 (2011). 
  

\bibitem{Suzuki:2012gi} 
  H.~Suzuki,
  arXiv:1209.2473 [hep-lat].; arXiv:1209.5155 [hep-lat].


\bibitem{Hori:2006dk} 
  K.~Hori and D.~Tong,
  JHEP {\bf 0705}, 079 (2007);  
%
\bibitem{Kanamori:2007yx} 
  I.~Kanamori, F.~Sugino and H.~Suzuki,
  Prog.\ Theor.\ Phys.\  {\bf 119}, 797 (2008); 
%
  I.~Kanamori,
  Phys.\ Rev.\ D {\bf 79}, 115015 (2009). 

\bibitem{Bergner:2007pu} 
  G.~Bergner, T.~Kaestner, S.~Uhlmann and A.~Wipf,
  Annals Phys.\  {\bf 323}, 946 (2008); 
 
\bibitem{Kadoh:2009sp} 
  D.~Kadoh and H.~Suzuki,
  Phys.\ Lett.\ B {\bf 684}, 167 (2010); 
  Phys.\ Lett.\ B {\bf 696}, 163 (2011). 
  
\bibitem{Kastner:2008zc} 
  T.~Kastner, G.~Bergner, S.~Uhlmann, A.~Wipf and C.~Wozar,
  Phys.\ Rev.\ D {\bf 78}, 095001 (2008); 
%
  M.~Heilmann, D.~F.~Litim, F.~Synatschke-Czerwonka and A.~Wipf,
  arXiv:1208.5389 [hep-th].
 
\bibitem{Kawai:2010yj} 
  H.~Kawai and Y.~Kikukawa,
  Phys.\ Rev.\ D {\bf 83}, 074502 (2011). 
\bibitem{Kamata:2011fr} 
  S.~Kamata and H.~Suzuki,
  Nucl.\ Phys.\ B {\bf 854}, 552 (2012). 
  
\bibitem{Schierenberg:2012pb} 
  S.~Schierenberg and F.~Bruckmann,
  arXiv:1210.5404 [hep-lat].

\bibitem{Wozar:2011gu} 
  C.~Wozar and A.~Wipf,
  Annals Phys.\  {\bf 327}, 774 (2012); 


\bibitem{Kim:2011cr} 
  S.~-W.~Kim, J.~Nishimura and A.~Tsuchiya,
  Phys.\ Rev.\ Lett.\  {\bf 108}, 011601 (2012). 
   
\bibitem{Steinacker:2010rh} 
  H.~Steinacker,
  Class.\ Quant.\ Grav.\  {\bf 27}, 133001 (2010). 

\bibitem{Chatzistavrakidis:2010tq} 
  A.~Chatzistavrakidis and G.~Zoupanos,
  SIGMA {\bf 6}, 063 (2010). 

   
  



  





  
  






%


\end{thebibliography}
\end{document}